\documentstyle[twocolumn,graphicx,aps,pra,floats]{revtex}
\begin{document}
\title  {Cost functions for pairwise data clustering ~\\ ~\\}
\author{Leonardo Angelini, Mario Pellicoro, Sebastiano Stramaglia} \address{Dipartimento Interateneo di Fisica, Universit\`{a} di Bari and
 I.N.F.N., Bari, Italy}
\author{ Luigi Nitti }\address{Dipartimento dell'Emergenza e dei Trapianti di Organi, Sezione Fisica Medica, Universit\`{a} di Bari
and  I.N.F.N., Bari, Italy} \maketitle
\begin{abstract}
Cost functions for non-hierarchical pairwise clustering are introduced, in the probabilistic autoencoder
framework, by the request of maximal average similarity between the input and the output of the autoencoder. The
partition provided by these cost functions identifies clusters with dense connected regions in data space;
differences and similarities with respect to a well known cost function for pairwise clustering are outlined.

\end{abstract}
\vspace{1.cm} Clustering methods aim at partitioning a set of
data-points in classes such that points that belong to the same
class are more alike than points that belong to different classes
\cite{ripley}. These classes are called clusters and their number
may be preassigned or can be a parameter to be determined by the
 algorithm. There exist applications of clustering in such diverse
fields as pattern recognition \cite{duda}, astrophysics
\cite{west}, communications \cite{linde}, biology \cite{alon},
business \cite{mantegna} and many others. Two main approaches to
clustering can be identified: parametric and non-parametric
clustering.

Non-parametric approaches make few assumptions about about the
data structure and, typically, follow some local criterion for the
construction of clusters. Typical examples of the non-parametric
approach are the agglomerative and divisive algorithms that
produce dendrograms. In the last years non-parametric clustering
algorithms have been introduced employing the statistical
properties of physical systems. The Super-Paramagnetic approach by
Domany and coworkers \cite{domany} exploits the analogy to a model
granular magnet: the spin-spin correlation of a Potts model,
living on the data-points lattice and with pair-couplings
decreasing with the distance, is used to partition points in
clusters. The synchronization properties of a system of coupled
chaotic maps are used in \cite{angelini} to produce hierarchical
clustering.

Parametric methods make some assumptions about the underlying data
structure. Generative mixture models \cite{bishop} treat
clustering as a problem of density estimation: data are viewed as
coming from a mixture of probability distributions, each
representing a different cluster, and the parameters of these
distributions are adjusted to achieve a good match with the
distribution of the input data. This can be obtained by maximizing
the data likelihood (ML) or the posterior (MAP) if additional
prior information on the parameters is available \cite{utsugi}.

Many parametric clustering methods are based on a cost function:
the best partition of points in clusters is assumed to be the one
with minimum cost. Often cost functions incorporate the loss of
information incurred by the clustering procedure when trying to
reconstruct the original data from the compressed cluster
representation: the most popular algorithm to optimize a cost
function is $K$-means \cite{bishop}. Starting from a statistical
{\it ansatz} and invoking maximum likelihood leads to a cost
function which has been observed to work for clustering financial
time series \cite{marsili}.

It is important to stress the difference between {\it central}
clustering, where it is assumed that each cluster can be
represented by a prototype \cite{rose}, and {\it pairwise}
clustering where data are indirectly characterized by pairwise
comparison instead of explicit coordinates \cite{hofman}; pairwise
algorithms require as input only the matrix of dissimilarities.
Obviously the choice of the measure of dissimilarity is not unique
and it is crucial for the performance of any pairwise clustering
method. It is worth remarking that it often happens that the
dissimilarity matrix violates the requirements of a distance
measure, i.e. the triangular inequality does not necessarily
holds.

Folded Markov chains are used in the Probabilistic Autoencoder
Framework to derive cost functions for clustering \cite{luttrel}.
Some examples of two-stage folded Markov chains, and the
corresponding algorithms for clustering and topographic mapping
\cite{bishop1}, are thoroughly analyzed in \cite{graepel}, where
it is also shown that the cost function for pairwise clustering,
introduced in \cite{hofman}, may be seen as a consequence of
Bayes' theorem and the requirement of minimal average distorsion
in a probabilistic autoencoder.

It is the purpose of this work to introduce a new class of cost
functions for pairwise clustering which can be obtained, in the
autoencoder frame, by requiring {\it maximal similarity} instead
of minimal distorsion. We show that the cost functions here
introduced provide a non-hierarchical clustering of points where
dense connected regions of points in the data space are recognized
as clusters.

Let us now discuss autoencoders described by one-stage folded
Markov chains. Let us consider a point $x$, in a data space,
sampled with probability distribution $P_0 \left(x\right)$; a code
index $\alpha \in \{1,\ldots,q\}$ is assigned to $x$ according to
conditional probabilities $P\left( \alpha |x\right)$. A
reconstructed version of the input, $x'$, is then obtained by use
of the Bayesian decoder:
\begin{equation}
P\left( x'|\alpha\right)={P\left( \alpha |x'\right) P_0
\left(x'\right)\over P\left( \alpha\right)}. \label{eq:bayes}
\end{equation}
The joint distribution of $x$, $x'$ and $\alpha$, describing this
encoding-decoding process, is
\begin{equation}
P\left( x,x',\alpha\right)=P_0 \left( x\right)P\left( \alpha
|x\right) P\left(x'|\alpha\right); \label{eq:bayes1}
\end{equation}
owing to (\ref{eq:bayes}), the joint distribution reads:
\begin{equation}
P\left( x,x',\alpha\right)={P_0 \left(x\right)P_0 \left(x'\right)
P\left( \alpha |x\right) P\left( \alpha |x'\right) \over P\left(
\alpha\right)}. \label{eq:bayes2}
\end{equation}
The conditional probabilities $\{ P\left( \alpha |x\right)\}$ are
the free parameters that must be adjusted to force the autoencoder
to emulate the identity map on the data space.

Let $d(x,x')$ be a measure of the distorsion between input and
output of the autoencoder. The average distorsion is then given
by:
\begin{equation}
{\cal D}=\sum_{\alpha=1}^q \int dx \int dx'{P_0 \left(x\right)P_0
\left(x'\right) P\left( \alpha |x\right) P\left( \alpha |x'\right)
\over P\left( \alpha\right)} d(x,x'). \label{eq:aved}
\end{equation}
Moreover, let s(x,x') be a measure of the similarity between input
and output; the average similarity is then given by
\begin{equation}
{\cal S}=\sum_{\alpha=1}^q \int dx \int dx'{P_0 \left(x\right)P_0
\left(x'\right) P\left( \alpha |x\right) P\left( \alpha |x'\right)
\over P\left( \alpha\right)} s(x,x'). \label{eq:aves}
\end{equation}
It is natural to postulate a one-to-one mapping between values of
distorsion and similarity, $s=F(d)$, with $F$ a strictly
decreasing function. A good autoencoder is obviously characterized
by a low value of ${\cal D}$ and high value of ${\cal S}$. However
we remark that the two requirements $Min({\cal D})$ and $Max({\cal
S})$, for reasonable choices of $F$, are not generically
equivalent.

Now we turn back to the clustering problem. Given a data-set
$\{x_i\}$ of cardinality $N$, partitioning these points in $q$
classes corresponds, in this frame, to design an autoencoder, with
$q$ code indexes, acting on data space. We choose the encoder to
be deterministic:
\begin{equation}
P\left( \alpha |x\right)=\delta_{\alpha\;\sigma(x)},
\label{eq:delta}
\end{equation}
$\sigma (x)\in \{1,\ldots,q\}$ being the code index associated to
$x$. The estimate for the average distorsion (\ref{eq:aved}),
based on the data-set at hand, is given by $\hat{{\cal D}}=N H_d
[\sigma]$, where we introduce the hamiltonian $H_d$ for the Potts
variables $\{\sigma_i\}$:
\begin{equation}
H_d [\sigma]=\sum_{\alpha=1}^q {\sum_{i,j=1}^N \delta_{\alpha
\sigma_i}\delta_{\alpha \sigma_j}d_{ij}\over \sum_{k=1}^N
\delta_{\alpha \sigma_k}}, \label{hamd}
\end{equation}
where $\sigma_i =\sigma (x_i)$, $d_{ij}=d(x_i,x_j)$. It turns out
that $H_d$ is equivalent to the cost function for pairwise
clustering, influential in the clustering literature, introduced
in \cite{hofman}.

The estimate for the average similarity is, similarly, given by
$\hat{{\cal S}}=-N H_s [\sigma]$, where we introduce the
hamiltonian $H_s$:
\begin{equation}
H_s [\sigma]=- \sum_{\alpha=1}^q {\sum_{i,j=1}^N \delta_{\alpha
\sigma_i}\delta_{\alpha \sigma_j}s_{ij}\over \sum_{k=1}^N
\delta_{\alpha \sigma_k}}. \label{hams}
\end{equation}

If we choose the autoencoder by minimizing the average distorsion,
then the best partition of the data-set in $q$ classes corresponds
to the ground state of $H_d$. If we choose it by maximizing the
average similarity, then the ground state of $H_s$ must be sought
for, instead. Since both $\{d_{ij}\}$ and $\{s_{ij}\}$ may be
taken positive, it follows that $H_d$ is characterized by
antiferromagnetic couplings between the Potts variables, while
$H_s$ is made of ferromagnetic couplings. Denominators in both
$H_d$ and $H_s$ serve to enforce the coherence among the $q$
clusters. In particular, without the denominator the ground state
of $H_s$ would correspond to a single big cluster.

The form of the function $F$, determining the relation between $s$
and $d$, has to be specified. In what follows we consider two
forms of this relation. A scale-free relation
\begin{equation}
s_{ij}=F_\gamma (d_{ij})=\left({d_{ij}\over \langle
d\rangle}\right) ^{-\gamma}, \label{dd}
\end{equation}
depending on the exponent $\gamma$, and a scale-dependent relation
\begin{equation}
s_{ij}=F_a (d_{ij})=\exp\left(-{1\over 2a^2}\left({d_{ij}\over
\langle d\rangle}\right)^2\right), \label{ff}
\end{equation}
dependent on the scale $a$. In the formulas above, $\langle d
\rangle$ is the average dissimilarity over all the pairs of
data-set points. The exponent $\gamma$ will be restricted to
assume small values so as to characterize the corresponding Potts
model by long-range ferromagnetic couplings; the scale parameter
$a$ will be bounded in $[0,1]$.

 At this point it is worth stressing that
minimization of the distorsion and maximization of the similarity
yield, in the autoencoder frame, different cost functions. The
hamiltonian $H_d$ embodies the requirement that pairs of distant
points (large $d_{ij}$) should belong to different clusters. On
the other hand, the hamiltonian $H_s$, for reasonable choices of
$F$, concentrates on pairs of close points (small $d$) and forces
them to belong to the same cluster. In other words, $H_s$ may be
seen to implement the idea that clusters should be searched for as
dense connected regions in the data space.

We describe now the application of the variational criterions for
clustering, described above, to some artificial and real
data-sets. We consider two optimization algorithms to find the
configuration of minimum cost: simulated annealing \cite{gelatt}
and mean-field annealing \cite{yuille}. Both approaches associate
a Gibbs probability distribution to the functional to be
optimized. Simulated annealing is a Monte-carlo technique which
samples the Gibbs distribution as the temperature is reduced to
zero, while mean-field annealing attempts to track an
approximation, to the mean of the distribution, known as {\it mean
field} approximation \cite{parisi}. We remark that an efficient
mean-field annealing algorithm for cost function (\ref{hamd}),
based on the EM scheme \cite{demp}, is described in \cite{hofman}:
the generalization of that algorithm to (\ref{hams}) is
straightforward.
\begin{figure}[ht]
\begin{center}
\includegraphics[width=8cm]{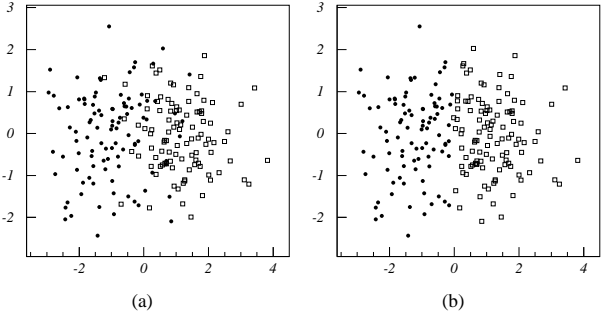}
\end{center}
\caption{(a) An artificial data set made of two Gaussian clusters,
each consisting of $100$ points. Empty squares and black circles
refer to the two different clusters. (b) Clustering result
obtained by minimization of $H_d$ (see the text).}
\end{figure}

In many cases cost functions $H_d$ and $H_s$ have very close
global minima. For example in Fig.1a we depict an artificial
data-set generated by two overlapping isotropic Gaussian
distributions. In this case the natural measure of dissimilarity
is Euclidean metrics, and we use $q=2$. In Fig.1b the
corresponding ground state of $H_d$ \cite{gr} is depicted: it is
very close to the Bayesian solution, i.e. the solution obtained
drawing the symmetry plane for the centers of the two Gaussians. A
similar partition is obtained minimizing, by simulated annealing,
$H_s$. As a measure of the difference between two partitions
$\{\sigma_i\}$ and $\{\eta_i\}$, we evaluate the following
quantity:
\begin{equation}
\epsilon = {1\over N(N-1)}\sum_{i=1}^N \sum_{j=1,j\ne i}^N
\left(\delta_{\sigma_i \sigma_j}-\delta_{\eta_i \eta_j}\right)^2
\label{ep} \end{equation} which counts the number of pairs of
points upon which the two partitions disagree. Using the
scale-dependent $F_a$, we find the ground state of $H_s$ to differ
from those of $H_d$ by $\epsilon < 0.01$ varying $a$ in
$[0.05,1]$. Analogously, using the scale-free $F_\gamma$, with
$\gamma\in [0.1,1.5]$, we find $\epsilon < 0.02$ when we compare
the ground state of $H_s$ with those of $H_d$. Hence, on this data
set, the cost functions introduced above work similarly within
wide ranges of $\gamma$ and $a$ values.

We find a similar behaviour with respect to the famous IRIS data
of Anderson \cite{anderson}. This data set has often been used as
a standard for testing clustering algorithms: it consists of three
clusters (Virginica, Versicolor and Setosa) and there are $50$
objects in ${\mathbf R}^4$ per cluster. Two clusters (Verginica,
Versicolor) are very overlapping. The clustering result, with
$q=3$ and minimizing $H_d$, consists of three clusters of $61$,
$39$ and $50$ points respectively, with $90\%$ of correct
classification percentage. We obtain exactly the same partition by
minimizing $H_s$ using a scale-free $F$ (with $\gamma \in
[0.15,1.45]$), and using a scale-dependent $F$ (with $a\in
[0.25,1]$). For $a\in[0.1,0.25]$ we obtain, in the scale-dependent
case, a slightly different partition with clusters' sizes $58$,
$42$, $50$ and correct classification percentage $93.3\%$. These
results show that also in the IRIS case the pairwise clustering
procedures by distorsion minimization and similarity maximization
are almost equivalent.

\begin{figure}[ht]
\begin{center}
\includegraphics[width=8cm]{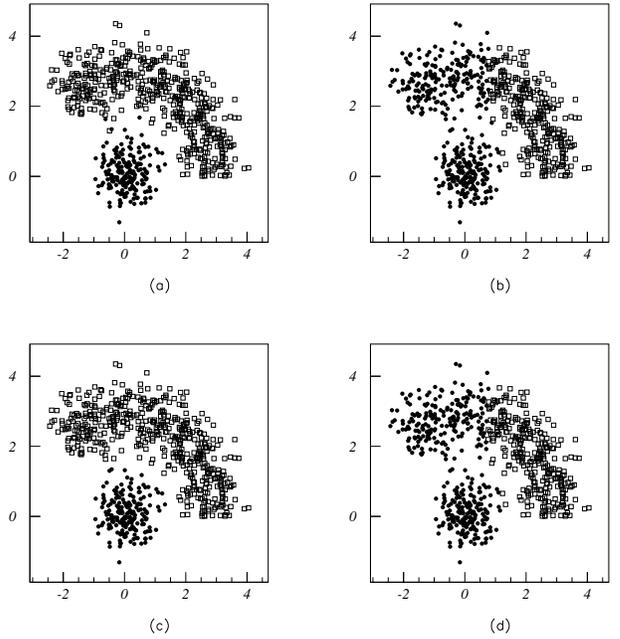}
\end{center}
\caption{(a) An artificial data set made of an elongated cluster
of $500$ points (empty circles) and a circular cluster of $200$
points (black circles). (b) Partition by minimizing $H_d$. (c)
Partition by minimizing scale dependent $H_s$ with $a<0.7$. (d)
Partition by minimizing scale dependent $H_s$ with $a>0.7$.}
\end{figure}
A typical situation resulting in different answers from $H_d$ and
$H_s$ is depicted in Fig.2a. This two-dimensional data-set is made
of an elongated cluster and a Gaussian distributed circular one.
It is evident that two dense connected regions are present, and
that the farthest pairs of points belong to the same connected
region. This is the type of data-set such that minimizing the
distorsion is not equivalent to maximizing the similarity. In
fig.2b the partition we obtain minimizing $H_d$ is depicted: it
fails to recognize the structure in the data-set. Let us now
consider the ground state of $H_s$ with the scale-dependent $F$.
For $a<0.7$ the ground state, depicted in Fig.2c, recognizes with
$99\%$ accuracy the data structure. At $a\sim 0.7$ a transition
phenomenon occurs: the configuration depicted in Fig.2c ceases to
be the global minimum, the new ground state (Fig.2d) being very
close to the solution by $H_d$.
\begin{figure}[ht]
\begin{center}
\includegraphics[width=8cm]{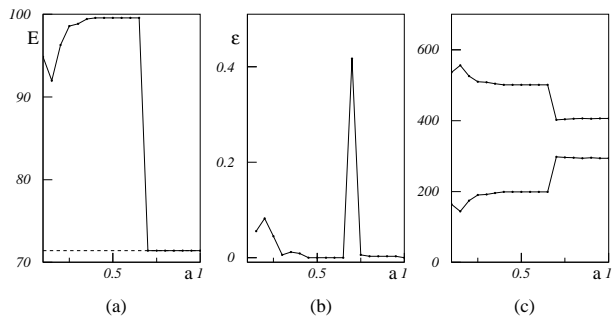}
\end{center}
\caption{(a) The efficiency (percentage of correctly classified
points) versus $a$, obtained on the data-set depicted in Fig.2 by
minimizing $H_s$ with scale-dependent $F$. The dashed line is the
efficiency obtained by minimization of $H_d$. (b) The $\epsilon$
parameter, (see the text) between partitions corresponding to
adjacent values of $a$, is plotted versus $a$ . (c) The size of
the two output clusters versus $a$.}
\end{figure}

In Fig.3a we depict the efficiency of the classification versus
the resolution parameter $a$, for the scale dependent $F$, while
in Fig.3b we consider a sequence of $a$-values and we plot the
$\epsilon$ between partitions corresponding to adjacent values of
$a$. The peak at $a=0.7$ is the indicator of the transition
between global minima. Finally, in Fig. 3c the size of the two
clusters, versus $a$, is depicted. Concerning the scale-free $F$,
in Fig.4 the same plots as in Fig.3 are depicted, showing that the
{\it good} minimum is stable for a wide range of $\gamma$.
\begin{figure}[ht]
\begin{center}
\includegraphics[width=8cm]{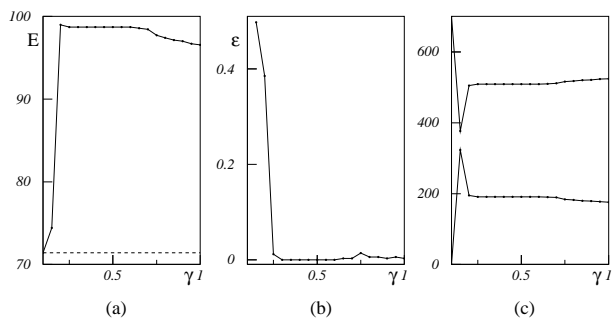}
\end{center}
\caption{(a) The efficiency (percentage of correctly classified
points) versus $\gamma$, obtained on the data-set depicted in
Fig.2 by minimizing $H_s$ with scale-independent $F$. The dashed
line is the efficiency obtained by minimization of $H_d$. (b) The
$\epsilon$ parameter, (see the text) between partitions
corresponding to adjacent values of $\gamma$, is plotted versus
$\gamma$ . (c) The size of the two output clusters versus
$\gamma$. }
\end{figure}
The choice of the optimization algorithm deserves a comment. All
the results described above are obtained by simulated annealing;
we also apply the mean-field annealing scheme, described in
\cite{hofman}, and we always find a configuration very close to
the one from simulated annealing, while spending less
computational time. This confirms that optimization algorithms
rooted on mean-field theory yield quickly a good solution on these
problems \cite{yuille}.

In summary, we address non-hierarchical pairwise clustering and,
working in the probabilistic autoencoder frame, we introduce a
class of cost functions arising from the request of maximal
average similarity between the input and the output of the
autoencoder. Our simulations show that the partition provided by
these new cost functions corresponds to extract dense connected
regions in data space, and that a relevant discrepancy with the
partition provided by the cost function introduced in
\cite{hofman} is to be expected in case of non-trivial geometry of
clusters. We note that the approach to clustering here described
has some similarities with the method in \cite{domany}: indeed in
both cases clustering is mapped onto a ferromagnetic Potts model
with couplings decreasing with the distance. In the
superparamagnetic approach, however, $q$ is not related to the
number of classes present in the data-set and one obtains
hierarchical clustering as the temperature of the Potts model is
varied. In the present case $q$ is the number of classes, which is
supposed to be known (non-hierarchical clustering), and the
denominators in the hamiltonian, ensuring clusters's coherence,
leads to a non-trivial ground state which reflects data structure.
We consider two classes of cost function. Scale-free cost
functions depend on the exponent $\gamma$, while scale-dependent
ones depend on the scale-parameter $a$. Varying $a$, i.e. changing
the resolution at which the data-set is processed, may give rise
to transitions between different partitions; in the scale-free
case, the clustering output is fairly stable, with respect to
$\gamma$, in a wide range.

Further work will be devoted to test these new cost functions on
other real applications and to study related issues, such as the
introduction of an {\it adaptive} relation between distorsion and
similarity, i.e. the function $s=F(d)$ might be depending on the
properties of the data-set in a neighbourhood of the pair of
points under consideration. It will be also important to develop
cluster-validity criterions to provide a means to choose an
optimal $q$ value in situations where the number of classes is
ambiguous.

\end{document}